\documentclass[journal]{IEEEtran}

\ifCLASSINFOpdf
\else
   \usepackage[dvips]{graphicx}
\fi
\usepackage{url}
\usepackage{amsmath}
\usepackage{graphicx}
\usepackage{placeins}
\usepackage{float}
\usepackage{bbm}
\usepackage{subfigure}
\usepackage{verbatim}
\usepackage{amssymb}
\usepackage{amsmath}
\usepackage{amsthm}
\usepackage{mwe}
\usepackage{docmute}
\usepackage[ruled, vlined]{algorithm2e}

\usepackage[disable]{todonotes}

\usepackage{graphicx}

\begin{document}

\title{Timely Target Tracking in Cognitive Radar Networks}
\author{William W. Howard, Charles E. Thornton, R. Michael Buehrer
\thanks{W.W. Howard, C.E. Thornton and R.M. Buehrer are with Wireless@VT, Bradley Department of ECE, Virginia Tech, Blacksburg, VA, 24061. \\
Correspondence:$\{$wwhoward$\}$@vt.edu  }
}

\maketitle
\pagenumbering{gobble}

\begin{abstract}
    We consider a scenario where a fusion center must decide which updates to receive during each update period in a communication-limited cognitive radar network. 
    When each radar node in the network only is able to obtain noisy state measurements for a subset of the targets, this means that the fusion center may not receive updates on every target during each update period. 
    If the set of nodes which are available to give updates in each update period is limited to the nodes with interesting updates, the problem is further constrained. 
    The solution for the selection problem at the fusion center is then non-stationary in time, and is not well suited for sequential learning frameworks where rewards have high temporal correlation. 
    The important parameters become the age of the most recent update for every track, and the measurement quality each node provides. 
    We derive an Age of Information-inspired track sensitive metric to inform node selection in such a network and compare it against less-informed techniques such as a multi-armed-bandit and random selection.
\end{abstract}


\section{Introduction}
Cognitive radar networks (CRNs) aggregate target information from dispersed nodes using a fusion center (FC) to create actionable information. 
This process relies on the timeliness of the target track information; when observations from radar nodes are delayed in arrival at the FC, the ability of an operator to make timely decisions is impeded. 
In a network with unlimited communications bandwidth, this may not be a problem, since the radar nodes can use whatever resources necessary to convey the observations. 
However, unlimited bandwidth is impractical. 
As the network size increases, the share of the limited communication resources allocated to each radar node will be reduced. 
This is even more apparent in sub-6 GHz radar networks. 
Spectrum comes at a premium, and must often be shared with primary or secondary users. 
This work address the timeliness problem for target track updating in CRNs by introducing an Age of Information (AoI) metric for node selection and comparing against several alternatives.

We model this fundamental constraint on network communications as a limit on the number of nodes which may provide updates in each update period and assume that a network with $M$ radar nodes may provide $C<M$ updates per update period.

The CRN we discuss in this work is composed of a single cognitive FC which coordinates updates and serves as an information aggregator to inform an operator. 
We measure the performance of a FC by analyzing the error of the target tracks it maintains. 
The CRN also contains $M$ independent radar nodes. 
Cognitive radar networks, as defined by Haykin \cite{haykin2006} and other works \cite{Martone_CRN_loop}, fall into one of two categories. 
Either the radar nodes have cognitive abilities, or the cognition in the network is limited to the fusion center. 
The network discussed in this work is of the second kind. 
The FC is able to monitor the environment and modify operating parameters to improve performance. 
Therefore, the radar nodes are not cognitive, as is common in related works \cite{howard2022_MMABjournal}, \cite{thornton2022_universaljournal}. 
Instead, they operate within some fixed frequency allocation and collect observations on those targets which are observable. 
One advantage to this structure is that traditional radars can be used while still providing some cognitive capability.

The environment contains $N_n$ low-altitude targets, such as UAVs, with $N_n$ possibly greater than $M$. 
The targets can enter or exit the environment from anywhere within the considered region, and can alter their speed and direction. 
Each radar node maintains a track for each observable target, and is capable of providing Kalman-filtered target state estimates to the FC upon request. 
In addition, each radar node is able to indicate to the FC when one of its tracks has exhibited ``interesting'' behavior - e.g., when a track initiates, retires, or alters velocity.

The FC operates on a periodic schedule. 
However, the network does not have enough capacity to provide target updates from every node during every update period. 
So, once per update period, the FC polls the nodes to gather information on availability. 
It then decides which subset of nodes from which to collect updates. 
Specifically, one update from a given node contains predicted states for all observable targets at that node. 
Since the FC receives several of these per update period, it fuses this information to update its own internal estimates of the global target state.

\paragraph{Contributions} 
This problem has not yet been explicitly addressed in the literature, but resembles scheduling problems where a central server must collect information from distributed nodes. 
As such, we borrow from the Age of Information literature to propose a node selection metric which is track age sensitive. 
To the best of our knowledge, this work represents the first consideration of Age-of-Information metrics in cognitive radar networks. 
In particular, we provide the first radar-inspired AoI metric, and show that it minimizes both the age of updates and the average track error, as compared to several alternative techniques.

\paragraph{Notation} We use the following notation. 
Matrices and vectors are denoted as bold upper $\mathbf{X}$ or lower $\mathbf{x}$ case letters.
Functions are shown as plain letters $F$ or $f$. 
Sets $\mathcal{A}$ are shown as script letters. 
The cardinality $|\mathcal{A}|$ of a set $\mathcal{A}$ refers to the number of elements in that set. 
The Euclidean norm of a vector $\mathbf{x}$ is written as $||\mathbf{x}||$. 
The time index of a FC update period is shown as $t$, while the time index of a radar node Coherent Pulse Interval (CPI) is given as $n$. 

\paragraph{Organization} 
Related work is discussed in Section II. Section III discusses the structure of the network in this paper, and Section IV covers our proposed techniques. Section V provides simulations to support our conclusions, which are in Section VI. 



\section{Background}
Cognitive radar networks are typically composed of several independent radar nodes \cite{howard2022_MMABjournal}, and occasionally use a central coordinator to provide cognitive feedback to the nodes \cite{howard2022_decentralizedconf}. 
``Independent'' means that the nodes are not controlled by a coordinator. 
The network in our current work adopts the fully-centralized variant of CRNs, as defined by Haykin \cite{haykin2005}. 
Cognition is considered here to be the ability to monitor the environment and modify operating parameters. 
In particular, we assume that the cognitive capability in the network is limited to the FC. The radar nodes simply collect target observations and do not modify their operating parameters.

The modifiable operating parameter available to the cognitive process in this work is the subset of nodes selected to provide updates in each update period. 
As the fusion center gains information about the underlying environment, the FC must learn which nodes are expected to have high-quality observations.

Age of Information metrics are popular tools for ensuring information freshness in a variety of applications. 
AoI was first proposed in \cite{AoI_initial}, and has gained considerable traction recently. 
The survey by Yates et al. \cite{AoI_survey2} covers recent contributions and applications and characterizes AoI as ``performance metrics that describe the timeliness of a monitor's knowledge of an entity or process.''

AoI has been used particularly often in federated learning problems \cite{AoI_scheduling}, \cite{AoI_timely}, where a central parameter server attempts to train a large machine learning (ML) model using numerous independent clients. 
Federated learning is important in domains which must respect data privacy, requiring a ML model to be trained in such a distributed fashion. 
AoI is useful in this field to ensure the global model is updated based on the most recent data, while maintaining the privacy of that data. 
Our current work does not have the same purpose; information freshness remains important to a CRN to ensure accurate information is presented to operators, but there is no condition on data privacy. 
Therefore we do not consider federated learning techniques.

AoI is also frequently applied in distributed sensor networks. 
In \cite{AoI_Dhillon}, the authors describe an uncrewed aerial vehicle (UAV) assisted IoT network which utilizes an AoI metric to to minimize information freshness. 
In this and in similar works \cite{AoI_sensor}, \cite{AoI_multiple}, a scheduler must assign resources to each of several nodes. 

A common metric in AoI problems is the peak age - the worst-case AoI. 
Let the age process be denoted as $\Delta(t)$. 
Assuming a unit-rate age process, the peak age of information (PAoI) is given as (\ref{eq:PAoI}) where there are $N(\tau)$ updates before $t=\tau$, and $A_n$ is the process age at the $n^{th}$ update \cite{peak_age_of_information}.  
A unit-rate process is one where the age increases by one in each time step. 

\begin{equation}
    \label{eq:PAoI}
    \Delta^{(p)}(t) = \lim_{\tau \to \infty} \frac{1}{N(\tau)} \sum_{n=1}^{N(\tau)} A_n = \mathbb{E}\left[A_n\right]
\end{equation}

\section{Network Structure}
We consider a radar network composed of multiple independent nodes, a single FC, and limited communication bandwidth. 
The scenario contains $N_n$ radar targets in each CPI $n$, where $N_n$ is time-varying. 
In each time step, the targets\footnote{Target state transition probabilities are constant between targets. In future work, we will investigate targets with dissimilar state transition probabilities and the implementation of a target maneuverability index. }: 
\begin{enumerate}
    \item Move through the scene according to their previous velocity. 
    \item Modify their velocity with probability $p_v$.
    \item Retire from the scene with probability $p_r$. 
\end{enumerate}
In addition, a set $\mathcal{N}_{new}$ of new targets enter the scene every time step with size specified by a Poisson distribution with parameter $p_s$, as in Eq. (\ref{eq:poisson}). 
\begin{equation}
\label{eq:poisson}
    Pr\left(|\mathcal{N}_{new}[n]|=k ; p_s \right) = \frac{1}{k!} \lambda^k e^{-p_s}
\end{equation}
It is the goal of each radar node to maintain a filtered track for each target in the environment. 
However, due to irregularities in the environment (i.e. interference, clutter, terrain, etc), each node is able to observe each target with probability $p_o$. 
When a new target enters the environment, each radar node $k \in \mathcal{N}$ adds it to the set $\mathcal{N}_k[t]$ with probability $p_o$. 
If this does not occur, the node is instead added to the set $\widetilde{\mathcal{N}}_k$\footnote{While the set $\mathcal{N}_k$ is observable by node $k$, the set $\widetilde{\mathcal{N}}_k$ is purely notational. }. 
Note that this only occurs when targets enter the environment, and the result persists until the target exits. 
Once a target has been added to $\mathcal{N}_k[t]$, it persists in $\mathcal{N}_k[t+\tau]$ until radar node $k$ fails to observe the target for $\tau$ consecutive update periods, at which point the retired target is removed from $\mathcal{N}_k$.

In addition, each radar node has a different observation quality for each target, due to differences in look angle, range, clutter, and other environmental factors. 
Specifically, the localization measurement variance for each node is drawn from an inverse-Gamma distribution (chosen as the Gaussian conjugate prior) as in Eq. (\ref{eq:invgamma}). 
\begin{equation}
\label{eq:invgamma}
    Pr(\sigma_{k, j} = \gamma ; a, b) = \frac{b^a}{\Gamma(a)}\left(\frac{1}{\gamma}\right)^{a+1} e^{\frac{-b}{\gamma}}
\end{equation}

Since the target tracking error at the FC will increase greatly when targets deviate from FC tracks, it is important for the radar nodes to signal to the FC when they believe a target has done something ``interesting''. 
Interesting behavior occurs when targets enter and exit the environment or change velocity. 
We measure this by evaluating the innovation in the Kalman filter for each target track. 
If the distance between the filter's predicted location and the observed location is greater than a threshold $d_I$ for any active target, the node raises a flag.

It is assumed that the independent radar nodes conduct observations in a pre-configured, non-interfering manner in their fixed spectrum allocations on an asynchronous basis. 
In other words, there is no assumption on pulse- or CPI-level synchronization of the radar nodes. 
Instead, once per update period, the FC polls each node to check whether it has any interesting observations (defined on the criteria above). 
Those nodes with flags raised are added to the set of available nodes $\mathcal{A}[t]$. 
Generally, $|\mathcal{A}[t]|$ is expected to be greater than the capacity $C$. 
In practice, we implement a penalty for selecting any node $k_n \notin \mathcal{A}[t]$.

In order to provide the most up-to-date information to an operator, the fusion center must optimize which nodes provide updates in each update period. 

A key metric which informs this error is the \emph{freshness} of the track information. 
As targets maneuver through the environment, target tracks which have not been updated recently will tend to drift away from their true value. 
We measure the freshness of the target track as the time since it was most recently updated, and denote the age for track $j$ in update period $t$ as $\Delta_j[t]$.

The second metric which the FC can use to inform node selection is \emph{measurement variance per track}. 
This quantity is derived at the radar node using the Kalman filter covariance for each active track. 
In each update period where node $k$ is selected, this variance is provided for each active track. 
The localization measurement variance for target $j$ being observed by node $k$ is written as $\sigma_{k,j}$ and is inverse-Gamma distributed as in Eq. (\ref{eq:invgamma}). 

The FC collects updates from each node in \emph{update periods}, which occur on random intervals. In simulation, update periods have a chance $P_{u}$ of occurring each CPI. 
An update period $t$ begins with a polling process, where the FC checks the update flag $A_k$ for each node $k$ to builds the node availability set $\mathcal{A}[t]$. 
Based on this information and using a node selection strategy, the FC then selects nodes $\mathcal{K} = [k_1, k_2, \dots, k_C]$ to provide updates.


\section{Track-Sensitive AoI Node Selection}
Broadly, AoI metrics track the time since certain quantities were updated, in order to optimize some criteria. 
In our scenario, the major quantities we are able to track are measurement variance (obtained from the individual node filter variance), and the time since each track at the FC has been updated (track freshness).

We form our objective function as Eq. (\ref{eq:obj}). 
In each update period, the FC selects nodes $\mathcal{K} = [k_1, k_2, ..., k_C]$ from which to collect updates. 
The set $\mathcal{A}[t]$ denotes node availability, and $\sigma_{k,j}$ is the localization variance experienced by node $k$ tracking target $j$.

The cases for $F_{k,j}$ denote the scenarios where the FC knows that radar node $k$ can see target $j$, can't see target $j$, and when the FC doesn't yet know. 
The term $\beta$, which is the result when the FC does not know whether radar node $k$ can see target $j$, represents an exploration factor, encouraging the FC to visit nodes which have not provided updates recently. 
The term $\gamma$ represent a penalty for each target that node $k_j$ cannot observe. 
The set $\hat{\mathcal{N}}$ contains all of the \emph{currently active target tracks} at the FC and is the FC's estimate of $\mathcal{N}$, the set of currently active targets. 
Finally, the term $\alpha$ provides a discounted reward for selecting nodes which does not have an interesting update. 
This generally ensures that when $|\mathcal{A}[t]|<C$, all nodes with interesting updates are selected before the remaining nodes are considered. 
Note that this metric is a joint optimization over the PAoI (due to the selection of the maximum-age tracks) and over observation variance. 

\begin{align}
\label{eq:obj}
    \mathcal{K} =& \max_{\mathcal{K} \in \mathcal{M}} \sum_{j \in \hat{\mathcal{N}}} \widetilde{\alpha}_k F_{k, j}\\
    \text{s.t. } & |\mathcal{K}| = C\\
    & F_{k,j} = \begin{cases}
    \Delta_j[t] \sigma_{k,j}^{-1} , & j \in \mathcal{N}_{k,FC}\\
    \gamma , & j \in \widetilde{\mathcal{N}}_{k,FC}\\
    \beta , & else
    \end{cases}\\
    & \widetilde{\alpha}_k = \begin{cases}
    1, & k \in \mathcal{A}[t]\\
    \alpha, & else
    \end{cases}
\end{align}

A naive approach may pick all nodes that observe the maximum-age track; this would cause that single track to receive many updates, but ignore the remaining tracks, and place no weight on the observation variance. 
Instead, this metric selects the minimum-variance estimate of the oldest tracks. 
This metric requires that $|\mathcal{K}|=C$, in order to maximize utilization of the communication resource.

In order to solve this optimization problem, we treat it as a bipartite matching problem and select the $C$ nodes which maximize the reward (\ref{eq:obj}). 

\vspace{0.1in}
\begin{algorithm}
\SetAlgoLined
 Receive $\hat{x}_j$ for all observable targets. \\
 Assign observations to existing tracks. \\ 
 Add new observations to $\mathcal{N}_{new}[n]$. \\ 
 Add retired tracks to $\mathcal{N}_{ret.}[n]$. \\ 
 Update $\mathcal{N}_k[n]$ as
    \begin{equation}
        \mathcal{N}_k[n] = (\mathcal{N}_k[n-1] \cup \mathcal{N}_{new}[n]) \backslash \mathcal{N}_{ret.}[n]
    \end{equation}\\
 Measure Kalman innovations per track (where $x^{(p)}_j$ is the predicted location for target $j$) as: 
    \begin{equation}
        I_j = ||\hat{x}_j - x^{(p)}_j||
    \end{equation}\\
 \uIf{$\exists j$ s.t. $I_j\geq d_I$}{   set $A_k=1$ and $a=1$. $\%$ Set flag}
 \uElseIf{$A_k=1$}{$a=a+1$ $\%$ Increment flag}
 \uIf{$a>a_{max}$}{$A_k=0$ $\%$ Reset flag}
 \caption{Actions for Radar Node $k$ in CPI $n$}
 \label{algo:node}
\end{algorithm}
\vspace{0.1in}
\vspace{0.1in}
\begin{algorithm}
\SetAlgoLined
 Select $\mathcal{K}[t]$ as Eq. (\ref{eq:obj}). \\
 Set $A_k = 0$ for all $k\in\mathcal{K}[t]$.  \\
 Receive $x^{(p)}_k[t]$ for all $k\in\mathcal{K}[t]$. \\
 Update tracks for targets $j$ observed by the selected nodes, where  
 \begin{equation}
     j \in \bigcup_{k\in\mathcal{K}[t]}\mathcal{N}_{k,FC}
 \end{equation}\\
  Begin new target tracks and retire tracks as appropriate. \\
 Reset track ages for updated tracks as
 \begin{equation}
     \Delta_j[t] = 1
 \end{equation}\\
 Increment track ages as 
 \begin{equation}
     \Delta_j[t] = \Delta_j[t-1] + 1, \;\; j \notin \bigcup_{\mathcal{K}[t]}\mathcal{N}_{k,FC}[t-1]
 \end{equation}\\

 \caption{Actions for Fusion Center in Update Period $t$}
 \label{algo:FC}
\end{algorithm}
\vspace{0.1in}

The FC maintains a table of which nodes can see which targets. 
When a node $k$ provides an update containing track information for target $j$, that target is added to the set $\mathcal{N}_{k,FC}$. 
A track $j$ at the FC is active until a node $k$ which had $j\in\mathcal{N}_{k,FC}[t]$ has $j\notin\mathcal{N}_{k, FC}[t]$. 
It is assumed that nodes retire tracks appropriately. 
In other words, the FC retires tracks as soon as updated by a radar node which has retired that track. 
The FC fuses radar observations from disparate nodes by taking a simple average of two observations reported simultaneously. 
The observations reported to the FC from the selected nodes are target variances and Kalman filter predictions evaluated when requested, rather than raw observations. 
Once the FC retires a target track, that target is no longer considered for this objective function.

The FC maintains an age $\Delta_j[t]$ for each track $j\in \hat{\mathcal{N}[t]}$. 
In update periods $t$ with $j \notin \bigcup_{k\in\mathcal{K}[t]}\mathcal{N}_{k,FC}[t-1]$, the age for track $j$ is incremented. 
In other words, when the FC does not expect an update for track $j$, the age is incremented. 
On the other hand, when a track is updated or initialized, the age is set to 1.

In summary, each radar node $k$ performs the actions in Algorithm \ref{algo:node} in each CPI, and the FC performs the actions in Algorithm \ref{algo:FC} in each update period.

\subsection{Alternative Node Selection Techniques}
We provide the following additional selection techniques for comparison. 
Multi-armed bandit models have been applied frequently to problems in cognitive radar, so we include an algorithm based on the Upper Confidence Bound (UCB) \cite{UCB_fischer}. 
We also discuss a random selection model. 

\subsubsection{UCB Node Selection}
The UCB is a metric used to balance the exploration-exploitation tradeoff in single-player bandit problems \cite{bandits}. 
The application to this problem is simple: over time, the FC selects $C$ ``arms'' and observes the corresponding average node measurement variances as rewards, which are used to inform future arm selections. 
We modify the traditional UCB algorithm slightly to provide support for multiple arm selection, and disabling arms corresponding to node availability. 
When fewer than $C$ nodes are available, we simply ensure the available nodes are selected, and choose the rest using UCB with all arms available.



\subsubsection{Random Node Selection}
Using random node selection, the FC would simply select nodes at random from the network. 
This represents the worst-case, least-informed performance. 
Obviously, worse performance could be obtained if particular knowledge of the node performance were available, but if the FC is completely uninformed, random selection represents the worst case. 
The performance of random node selection will still be in excess of any single node's performance, since information from multiple nodes is still being fused. 


\section{Results}
The following simulations include a scenario with a time-average of $\overline{N}=20$ active targets. 
To maintain a constant average number of targets, $p_s$ is set to $\overline{N}p_r$.
The CRN consists of a single FC using the track-sensitive AoI node selection algorithm unless specified otherwise. 
There are 15 radar nodes in the network. 
FC update periods occur with a probability of $P_u=0.25$ per CPI.  
Other parameters are specified in Table \ref{table:params}. 

\begin{table}
\centering
\begin{tabular}{||c | c ||} 
 \hline
 Parameter & Value \\ [0.5ex] 
 \hline\hline
 $p_s$ - Probability of new targets & 0.1 \\ 
 \hline
 $p_r$ - Probability of retiring & 0.005 \\
 \hline
 $p_t$ - Probability of turning & 0.01 \\
 \hline
 $p_o$ - Probability of observing each target & 0.2 \\
 \hline
 $a$ - Measurement variance & 2 \\
 \hline
 $b$  - Measurement variance & 1 \\
 \hline
 $\alpha$  - Discount for no node flag & 0.01 \\
 \hline
 $\beta$  - Exploration factor & 1 \\
 \hline
 $\gamma$  - Penalty for unobserved target & -1 \\
 \hline
\end{tabular}
\vspace{0.1in}
\caption{Simulation parameters. }
\label{table:params}
\end{table}

\begin{figure}
    \centering
    \includegraphics[scale=0.55]{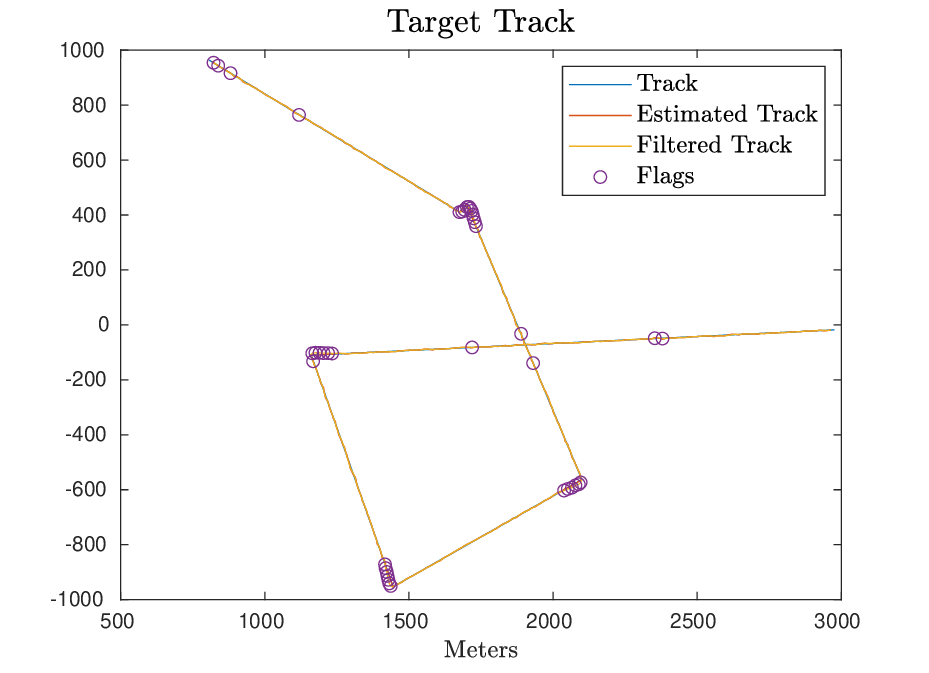}
    \caption{Single-node target track of relatively low variance. When the target changes velocity or exits the environment, the node raises a flag, visualized as a circle. }
    \label{fig:flags}
\end{figure}

Figure \ref{fig:flags} plots the ground track for a single target, and a single-node estimate of that track. 
We see that with low variance, the node is able to maintain an accurate track of the target. 
When the Kalman innovation for this track exceeds the threshold, the node indicates this with a flag, denoted with circles. 
We can see that the node accurately observes when the target turns and retires. 

\begin{figure}
    \centering
    \includegraphics[scale=0.55]{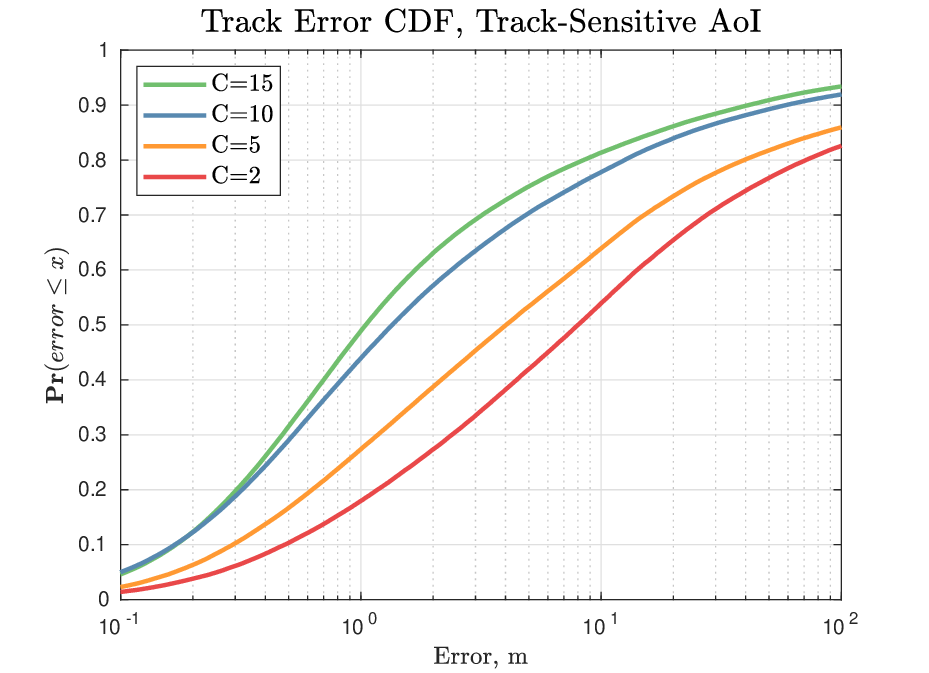}
    \caption{Time-averaged percent of active tracks which fall under a given error threshold. The simulated network contains 20 radar nodes. We see that only collecting feedback from 10 of the nodes leads to little loss in performance. Reducing the capacity further to 2 nodes results in reduced performance. }
    \label{fig:capacity_comp}
\end{figure}

The key metrics we wish to examine are radar tracking performance and FC track age. 
We measure radar tracking performance by inspecting the percentage of tracks which maintain error under a given threshold in each update period. 
This is equivalent to an empirical cumulative distribution function of the track error. 
Better performance corresponds to lower error, i.e. plots which shift towards the left. 
In Fig. \ref{fig:capacity_comp}, we see the tracking error for the network described above with the capacity set to 15, 10, 5, and 2. 

Note that using $C=15$ corresponds to full-feedback; updates are collected from every node in every update period. 
We can see that reducing $C$ by $33\%$ to 10 results in very slightly reduced performance, however going down to $C=5$ results in much worse performance. 
This can be attributed to the number of targets we are able to observe in each update: there is an average of 20 active targets, but with $C=2$ nodes we can observe at most $N * p_o * C = 8$ targets per update.

\begin{figure}
    \centering
    \includegraphics[scale=0.55]{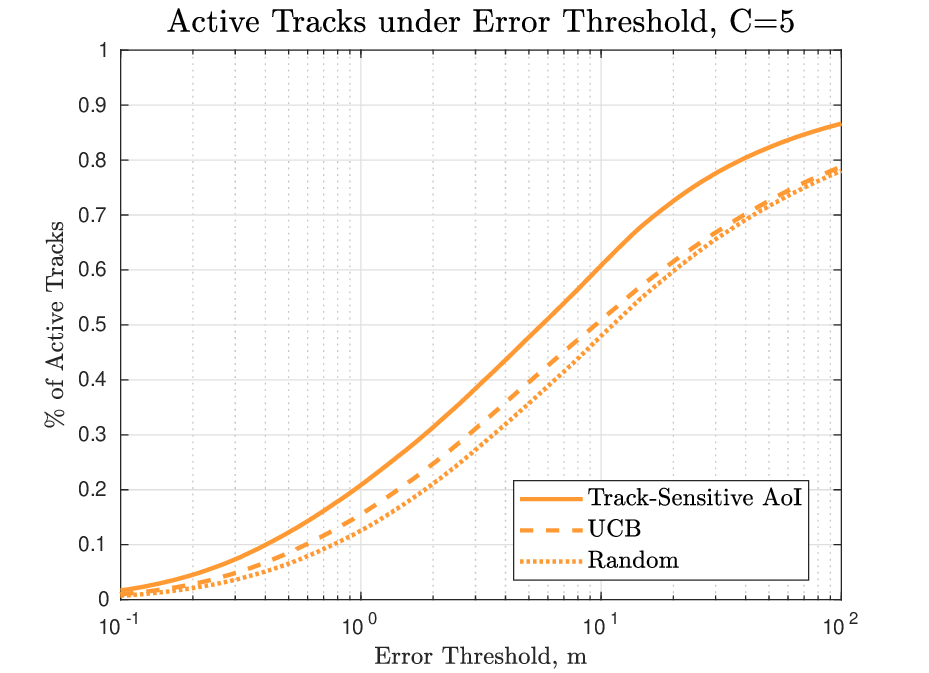}
    \caption{Time-averaged percent of active tracks which fall under a given error threshold for track sensitive AoI, UCB, and random node selection, for a capacity of 5. $10\% - 15\%$ more tracks fall under a given error threshold for AoI over random selection. UCB performs worse than random selection for high error cases due to greedy selection of low-variance nodes, which ignores high-variance tracks. }
    \label{fig:alg_comp}
\end{figure}

We also compare the tracking error between the different selection algorithms in Figure \ref{fig:alg_comp}. 
Since track sensitive AoI is able to take into account both the node variance and track age, it is able to perform the best. 
Random selection exhibits the worst performance, since it represents the least-informed selection algorithm.

\begin{figure}
    \centering
    \includegraphics[scale=0.55]{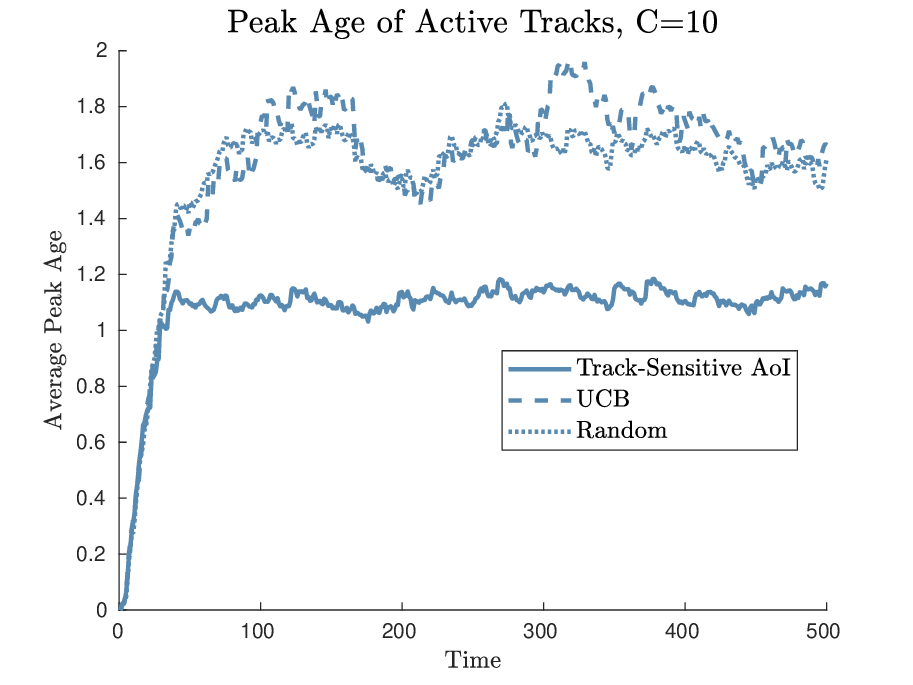}
    \caption{Peak age averaged over all active tracks for three different strategies. Since the AoI metric accounts for peak age, we see that it performs the best. Then, since UCB will be have an incentive to select nodes which see more targets, it performs better than random selection.  }
    \label{fig:peak_age}
\end{figure}

We are optimizing for track peak age as well as variance through the AoI metric. 
In Figure \ref{fig:peak_age} we see that the AoI metric outperforms the others in terms of peak age. 
This is because this technique is able to jointly optimize over track variance and track age, reducing the age of tracks upon updating. 

\begin{figure}
    \centering
    \includegraphics[scale=0.55]{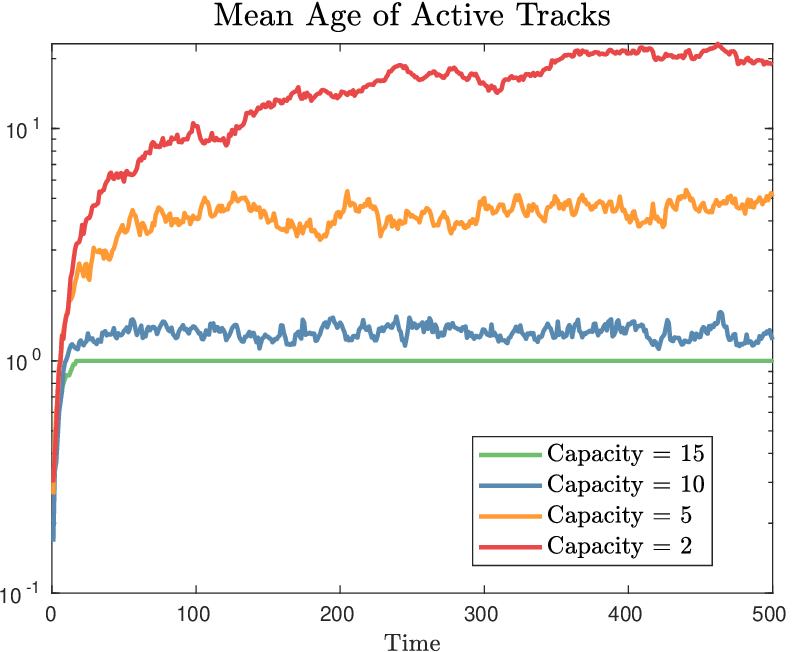}
    \caption{Mean age of each active target track at the FC. As the network capacity increases, the mean age of active targets reduces. A capacity of ten is sufficient to maintain an average age near one, while any lower capacity results in an average age of several time steps. }
    \label{fig:mean_age}
\end{figure}

Another way to visualize the capacity limitation is by examining the mean age of active tracks at the FC. 
In Fig. \ref{fig:mean_age}, we see that higher-capacity networks are able to maintain a lower mean age. 
This is simply because of the quantity of targets the FC can update. 

\begin{figure}
    \centering
    \includegraphics[scale=0.55]{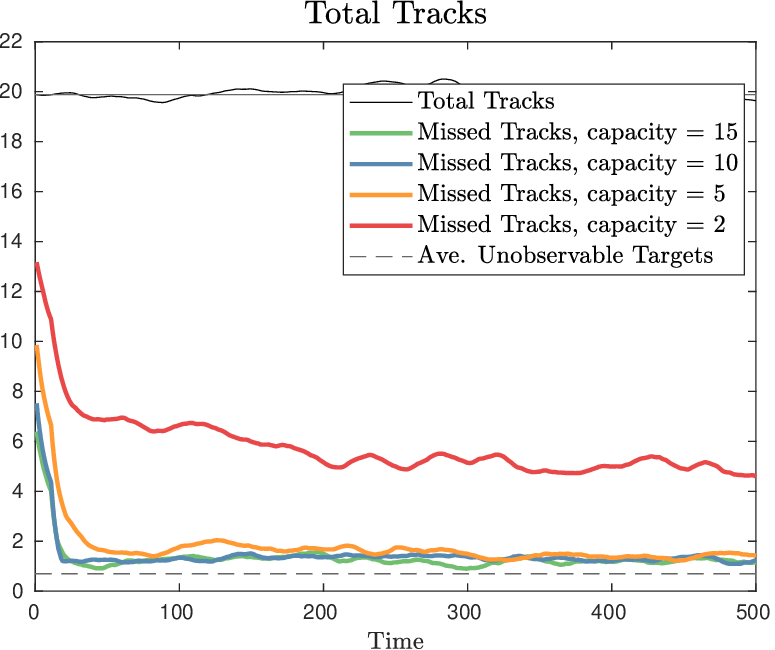}
    \caption{Number of targets averaged over many simulations. We can see that the network capacity is inversely correlated with the number of missed tracks. At the beginning of the simulation, the number of missed tracks spikes due to initialization. Even in the case of full feedback, there is some quantity of missed tracks due to nodes unobservable to the network. }
    \label{fig:total_tracks}
\end{figure}

Lastly, we can examine the number of targets which exist in the environment but do not have tracks at the FC. 
Missed tracks are caused by targets which are unobserved due to capacity limitations or observation limits. 
In Fig. \ref{fig:total_tracks}, we see both the total number of active tracks in each time step, and the number of missed tracks at the FC. 
In addition, due to the fixed probability per node of observing a given target $p_o$, there will be some number of targets which are unobservable to the network. 
Specifically, for observation probability $p_o$, there will be $\overline{N}(1-p_o)^M$ targets unobservable to the network. 
For $p_o=0.2$, $M=15$ nodes, and an average of $\overline{N}=20$ targets, this works out to $<1$ unobservable targets. 
Fig. \ref{fig:total_tracks} shows this as a dashed line. 

%

\section{Conclusions and Future Work}
In this work, we demonstrated the efficacy of a track sensitive AoI metric for node selection in communication-limited cognitive radar networks. 
We compared this approach against a multi-armed bandit model and a random selection algorithm. 
We showed that in both target tracking and information freshness, our proposed method outperforms the others we investigated.
This represents the first work in this field, drawing inspiration from the AoI literature. 

In a real system, this type of optimization could result in simpler track management and lower communication requirements. 
Due to the decreased PAoI, tracking performance should increase, especially for a large number of targets.

We intend to expand this work to include feedback for node control and explore further node-selection criteria. 
In addition, we will investigate more realistic target models with dissimilar state transition probabilities between targets. 
This will allow the FC to prioritize node updates which contain targets with high maneuverability indices. 
Future work will remove some simplifying assumptions on the radar signal processing; namely, this work assumes that node observation quality is constant in time and that nodes perfectly assign detections to target tracks.

\bibliographystyle{IEEEtran}
\bibliography{bibli}

    
\end{document}